\pdfoutput=1
\documentclass[aip,amsmath,amssymb, reprint,sort&compress,super,comma]{revtex4-1}

\usepackage{amsmath,amssymb}
\usepackage{graphicx}
\usepackage{color}
\usepackage{dcolumn}
\usepackage{bm}
\usepackage{subfigure}
\usepackage{hyperref}

\usepackage{placeins}
\usepackage{enumitem}

\newcommand*{\secref}[1]{Section~\ref{sec:#1}}

\newcommand*{\figref}[1]{Fig.~\ref{fig:#1}}
\renewcommand*{\eqref}[1]{Eq.~(\ref{eq:#1})}
\newcommand*{\eqsref}[3]{Eqs.~(\ref{eq:#1}), (\ref{eq:#2}), and (\ref{eq:#3})}

\newcommand{\citetdot}[1]{\citeauthor{#1}\citep{#1}}
\newcommand{\citetcomma}[1]{\citeauthor{#1},\citep{#1}}

\definecolor{background-color}{gray}{0.98}

\begin{document}

\title{Data-driven and constrained optimization of semi-local exchange and non-local correlation functionals for materials and surface chemistry}
\author{Kai Trepte}
\email{ktrepte@slac.stanford.edu}
\affiliation{\mbox{SUNCAT Center for Interface Science and Catalysis, SLAC National Accelerator Laboratory}\\ \mbox{Menlo Park, CA 94025, USA}}
\author{Johannes Voss}
\email{vossj@slac.stanford.edu}
\affiliation{\mbox{SUNCAT Center for Interface Science and Catalysis, SLAC National Accelerator Laboratory}\\ \mbox{Menlo Park, CA 94025, USA}}

\begin{abstract}
Reliable predictions of surface chemical reaction energetics require an accurate
description of both chemisorption and physisorption. Here, we present an empirical approach
to simultaneously optimize semi-local exchange and non-local correlation of a density
functional approximation to improve these energetics. A combination of
reference data for solid bulk, surface, and gas-phase chemistry and physical
exchange-correlation model constraints leads to the \mbox{VCML-rVV10}
exchange-correlation functional. Owing to the variety of training data,
the applicability of \mbox{VCML-rVV10} extends beyond surface chemistry simulations.
It provides optimized gas phase reaction energetics and an accurate description of
bulk lattice constants and elastic properties.
\end{abstract}

\maketitle

\begin{figure}[h]
\centering
\colorbox{background-color}{
\fbox{
\begin{minipage}{1.0\textwidth}
\includegraphics[width=50mm,height=50mm]{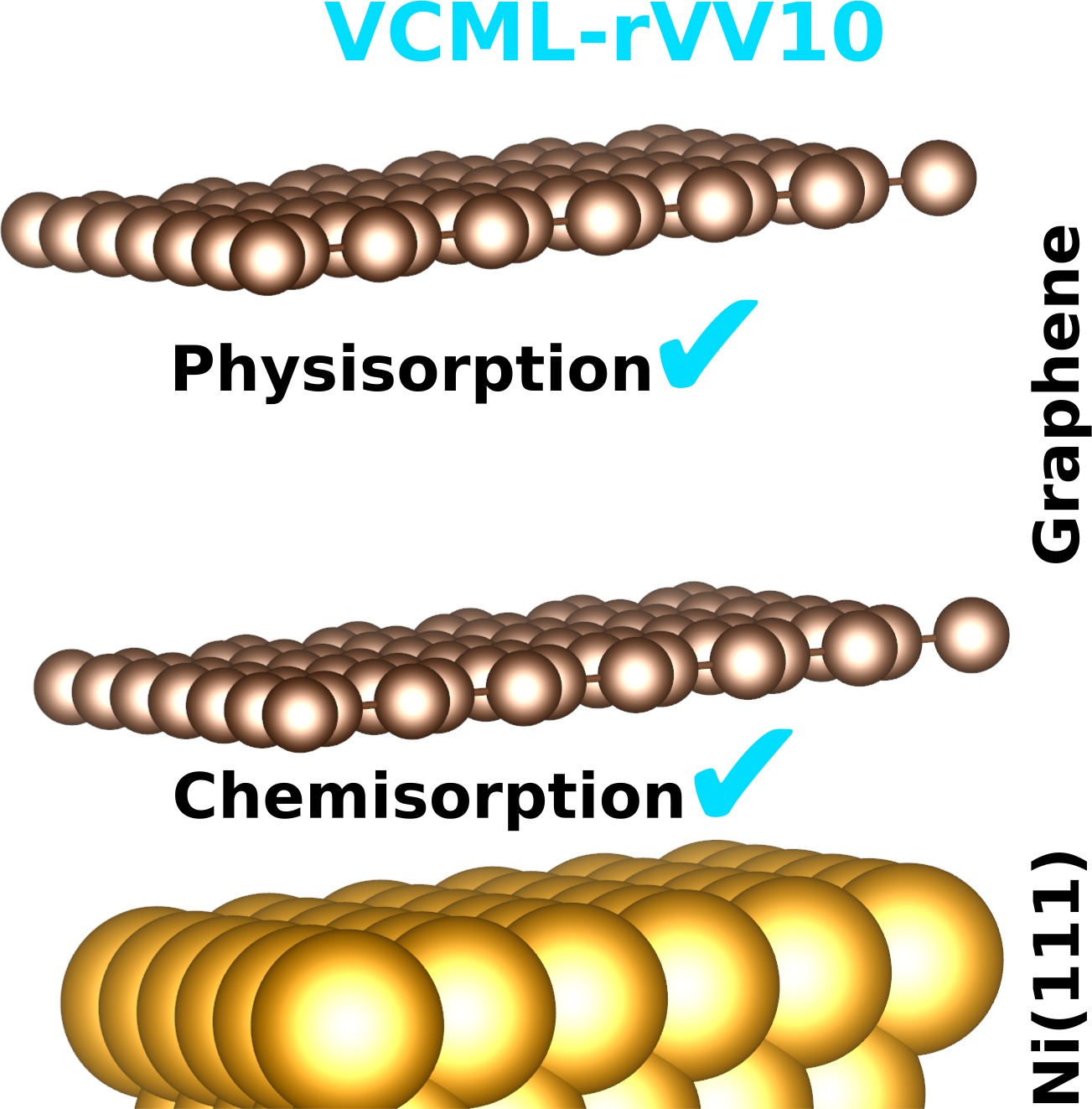} \\
VCML-rVV10 is a meta-GGA exchange-correlation functional which
has been derived from a simultaneous optimization of semi-local
exchange and non-local, i.e., van der Waals correlation. Training
the functional on a variety of physical and chemical data as well
as enforcing a number of constraints of the exact functional makes
VCML-rVV10 accurate for surface chemistry without compromising on the
prediction of bulk lattice constants and molecular interactions.
\end{minipage}
}}
\end{figure}

  \makeatletter
  \renewcommand\@biblabel[1]{#1.~}
  \makeatother

\bibliographystyle{unsrtnat} 

\renewcommand{\baselinestretch}{1.0}
\normalsize

\clearpage

\section{Introduction}
Kohn-Sham (KS) density functional theory (DFT)\cite{Hohenberg1964_B864,Kohn1965_A1133}
is one of the most widely used electronic structure theories. It provides reasonable accuracy for many physical properties
of atoms, molecules, surfaces, and solids.\cite{Kohn1996_12974}
For the practical use of KS DFT, approximations to electronic exchange and correlation (XC) are needed.
These so-called XC functionals have been
developed for decades.\cite{Burke2011_150901,Mardirossian2017_2315} There is a variety of both
empirical\cite{Becke1993_5648,Zhao2005_161103,Perverati2011_20120476,Mardirossian2014_9904} and non-empirical
functionals.\cite{Perdew1996_3865,Perdew2008_136406,Sun2015_036402,Garza2018_3083,Furness2020_8208}
In the latter approaches it was established that fulfilling analytical constraints\cite{Perdew2014_18A533}
leads to functionals with improved predictions of molecular and bulk properties.\cite{Sun2015_036402,Bokdam2017_145501,Zhang2018_063020}
Typically, empirical as well as non-empirical functionals describe some physical and chemical properties
with improved accuracy at the price of worse predictions of other properties.

To obtain an XC functional that accurately describes a wide range of physical properties,
strategies from the empirical and non-empirical XC functional development approaches can be combined.
Constraints from the latter can be imposed while fitting a multi-parameter empirical XC functional form
to experimental and quantum chemistry reference data.

There are several approaches combining constraints and data for XC functionals.\cite{Brown2021_2004,Sparrow2021__,Zhao2006_364}
Recently, our group has applied such a combined approach to construct a multi-purpose,
constrained and machine learned (MCML) XC functional. MCML shows improved predictions
of surface and gas phase reactions without sacrificing the good description of bulk
properties.\cite{Brown2021_2004}
The MCML functional is a so-called semi-local functional, lacking an explicit description
of non-local, {\it i.e.}, van der Waals (vdW) type correlation. Supplementing a semi-local
functional with vdW correlation does not only affect the performance for vdW-dominated interactions.
For example, some semi-local approximations tend to overbind chemisorbed systems.\cite{Duanmu2017_835,Sharada2019_035439}
Adding attractive dispersion forces can increase this tendency even further.
The semi-local part of a functional can be
optimized to compensate for effects of overbinding or lattice spacing contraction from
the attractive vdW terms.\cite{Wellendorf2012_235149,Klimes2009_02201} On the other hand, the vdW part
can be optimized using, {\it e.g.}, the parameterized, semi-empirical rVV10 non-local vdW
term.\cite{Vydrov2010_244103,Sabatini2013_041108,Peng2016_041005} With that,
performance on dispersion-dominated benchmark data is optimized while fixing the underlying semi-local functional.

\begin{figure}[h!!]
    \centering
    \includegraphics[width=0.49\textwidth]{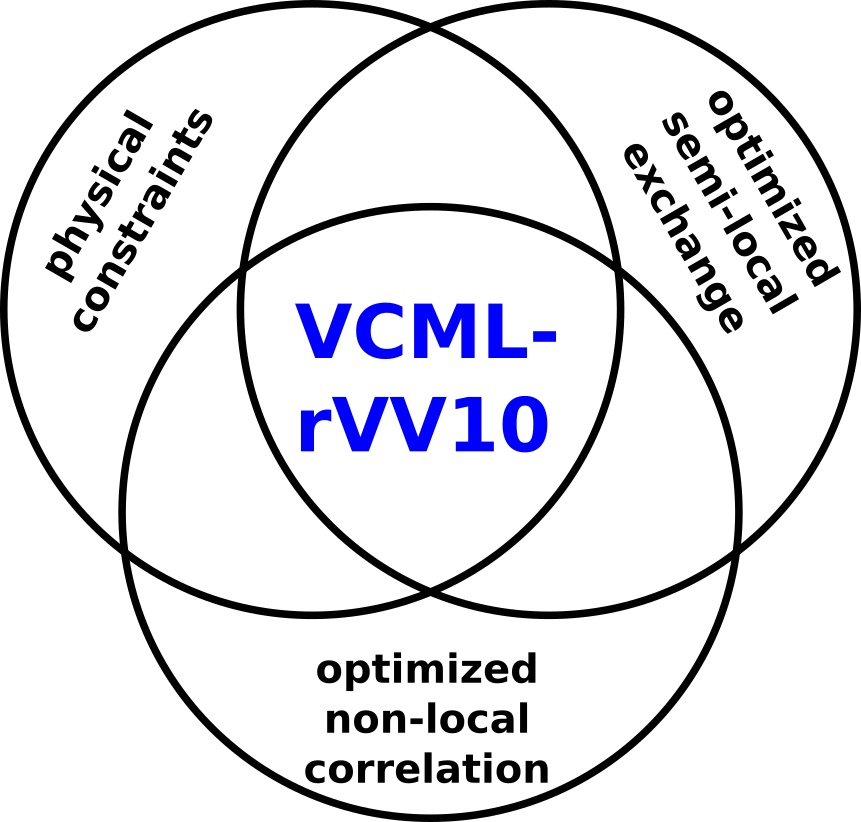}
    \caption{Basic outline of the \mbox{VCML-rVV10} optimization.}
    \label{fig:introVCML}
\end{figure}
In the present work, we optimize the semi-local exchange and non-local correlation parts of
a vdW XC functional simultaneously for several bulk, surface, and gas phase property predictions.
This simultaneous optimization of the semi-local exchange functional form and an
rVV10 non-local term enables us to construct a functional that is accurate for a range
of materials properties governed by different types of chemical bonds. There is no
compromise between the accuracy for the description of ionic, covalent or metallic bonds and
dispersive interactions, or vice versa. Incorporating physical constraints into this empirical
approach leads to the \mbox{VCML-rVV10} functional (vdW
functional using constraints and machine learning, employing
the rVV10 formalism), see \figref{introVCML}. This name shall clearly separate
our functional from MCML. The semi-local exchange functional parts of MCML and
\mbox{VCML(-rVV10)} are different, and the latter is not simply a vdW-supplemented form of MCML.
While we optimize the semi-local exchange as well
as the non-local correlation part of our functional,
for semi-local correlation we employ REGTPSS,\cite{Perdew2009_026403} in analogy to \citetdot{Brown2021_2004}

This article is structured as follows. In \secref{theory},
a brief introduction of the theoretical background including the rVV10 formalism is given.
The used computational parameters are shown in \secref{compdetails}.
All employed data sets are described in \secref{datasets}.
Thereafter, our functional optimization approach is presented in \secref{fitting}.
The main results are discussed in \secref{results}. A conclusion is presented in \secref{conclusion}.

\section{Theoretical background}
\label{sec:theory}
Atomic units are being used throughout this section.
In the KS formulation of DFT, the
total energy of a system with total electron density
\begin{equation}
    n(\textbf{r}) = n^{\uparrow}(\textbf{r}) + n^{\downarrow}(\textbf{r}),
\end{equation}
where $n^{\uparrow}$ and $n^{\downarrow}$
are the spin densities of a system with collinear spins,
can be expressed as
\begin{equation}
    E^{\text{KS}} = T_{\text{S}}[n^{\uparrow},n^{\downarrow}] + E_{\text{ext}}[n] + E_{\text{H}}[n] + E_{\text{XC}}[n^{\uparrow},n^{\downarrow}]. \label{eq:KS}
\end{equation}

Here, $T_{\text{S}}$ is the non-interacting kinetic
energy, $E_{\text{ext}}$ is the energy of the interaction
with an external potential, and $E_{\text{H}}$ is
the Hartree energy of the electrons. Exchange and
correlation effects are treated via the exchange-correlation
functional $E_{\text{XC}}$.

The electron density is
the fundamental building block of DFT. It
determines the energy as well as any
other property. While DFT is in principle exact, the $E_{\text{XC}}$
needs to be approximated.
Such approximations can be categorized according to their ingredients in the functional form.
This is known as the Jacob's ladder of DFT.\cite{Perdew2001_1}
The lowest rung is given by the local (spin) density approximation (LSDA).
Only the density itself is taken as a parameter.
Additionally including the density gradient characterizes
generalized gradient approximations (GGAs).
\mbox{Meta-GGAs} also include a dependence on the kinetic energy density
of the occupied orbitals.

The exchange energy for meta-GGAs is
defined as an integral of the exchange energy density per electron
of the homogeneous electron gas (HEG) at density $n(\textbf{r})$
\begin{equation}
 \epsilon^{\rm HEG}_{\text{X}}[n(\textbf{r})]=-\frac{3}{4}(3/\pi)^{1/3}n(\textbf{r})^{1/3},
\end{equation}
multiplied by an exchange enhancement factor $F_{\text{X}}$
\begin{equation}
    E_{\text{X}}[n(\textbf{r})] = \int \text{d}^3 r n(\textbf{r}) \epsilon^{\text{HEG}}_{\text{X}}[n(\textbf{r})] F_{\text{X}}(s(\textbf{r}),\alpha(\textbf{r})). \label{eq:Ex}
\end{equation}
Here,
\begin{equation}
    s(\textbf{r}) = \frac{|\boldsymbol{\nabla} n(\textbf{r})|}{2 k_{\text{F}}(\textbf{r}) n(\textbf{r})}
\end{equation}
is the reduced density gradient with the Fermi wave vector
\begin{equation}
    k_{\text{F}}(\textbf{r}) = \sqrt[3]{3 \pi^2 n(\textbf{r})}.
\end{equation}
The dependence
on the kinetic energy density can be described as
\begin{equation}
    \alpha(\textbf{r}) = \frac{\tau^\text{KS}(\textbf{r}) - \tau^\text{Weizs\"acker}(\textbf{r})}{\tau^{\text{HEG}}(\textbf{r})} \label{eq:alpha} ,
\end{equation}

with
\begin{align}
    \tau^\text{KS}(\textbf{r}) & = \frac{1}{2} \sum_i f_i |\boldsymbol{\nabla} \phi_i(\textbf{r})|^2, \\
    \tau^\text{Weizs\"acker}(\textbf{r}) & = |\boldsymbol{\nabla} n(\textbf{r})|^2 / (8 n(\textbf{r})), \\
    \tau^{\text{HEG}}(\textbf{r}) & = \frac{3}{10}(3 \pi^2)^{2/3} n(\textbf{r})^{5/3}
\end{align}

being the kinetic energy density based on the KS
orbitals $\phi_i(\textbf{r})$ with occupation $f_i$,
the von Weizs\"acker kinetic energy density
describing the single orbital limit, and
the non-interacting kinetic energy density
of a HEG at density $n(\textbf{r})$, respectively.
The term $\tau^\text{Weizs\"acker}(\textbf{r})$
vanishes for zero charge density gradient and equals
$\tau^{\text{KS}}(\textbf{r})$ for a single orbital
density. Accordingly, Eq.~(\ref{eq:alpha}) approaches
zero in the single orbital limit and one in the
homogeneous limit where $\boldsymbol{\nabla} n(\textbf{r}) = 0$
and $\tau^{\text{KS}}(\textbf{r}) = \tau^{\text{HEG}}(\textbf{r})$.
With that, the parameter $\alpha$ can characterize
the type of bonding in a system.

For spin-polarized systems, $E_\text{X}$ is obtained
via the spin scaling relation
\begin{equation}
    E_\text{X}[n^\uparrow(\textbf{r}),n^\downarrow(\textbf{r})] = \frac{E_\text{X}[2n^\uparrow(\textbf{r})] + E_\text{X}[2n^\downarrow(\textbf{r})]}{2}.
\end{equation}
The exchange-enhancement factor $F_{\text{X}}$ is optimized
in this work by training against reference data, see \secref{fitting} for more details.

In the rVV10\cite{Vydrov2010_244103,Sabatini2013_041108}
methodology, a non-local correlation energy is introduced as
\begin{equation}
    E_{\text{c,nl}} = \frac{1}{2} \int \int \text{d}^{3}r \text{d}^{3}r' n(\textbf{r}) \Theta(n(\textbf{r}),n(\textbf{r}')) n(\textbf{r}').
\end{equation}
The non-local kernel as described by \citet{Sabatini2013_041108}
is given by

\begin{widetext}
\begin{equation}
    \Theta(n(\textbf{r}),n(\textbf{r}'))^{\text{rVV10}} = -\frac{3}{2} \frac{1}{(q(\textbf{r})R^{2}+1)(q'(\textbf{r}')R^{2}+1)(q(\textbf{r})R^{2}+q'(\textbf{r}')R^{2}+2)},
\end{equation}
\end{widetext}
in which
\begin{equation}
    q(\textbf{r}) = \frac{\omega_{0}(n(\textbf{r}),|\boldsymbol{\nabla} n(\textbf{r})|)}{k(\textbf{r})}
\end{equation}
and $R = |\textbf{r} - \textbf{r}'|$. The same expression
is obtained for $q'$ by replacing $\textbf{r}$ with $\textbf{r}'$.
The term $\omega_{0}$ depends on the local band gap as
well as the plasma frequency. For more information, see \citet{Sabatini2013_041108}
as well as the supporting information (SI), Sec.~1. Importantly
for further discussion, the term
\begin{equation}
    k(\textbf{r}) = 3 \pi b \left(\frac{n(\textbf{r})}{9\pi}\right)^{\frac{1}{6}}
\end{equation}
determines the short-range damping of the $R^{-6}$ divergence
of the non-local kernel. The parameter $b$ needs to be
optimized for any functional $E_{\text{XC}}$ that
$E_{\text{c,nl}}$ is added to. Simply speaking, the
larger the $b$ parameter, the weaker the vdW interaction.
The optimization of this parameter is discussed
in \secref{fitting}. For additional details we refer to
the original works of \citet{Vydrov2010_244103} and \citetdot{Sabatini2013_041108}

\section{Computational details}
\label{sec:compdetails}
In this work, the Vienna ab initio Simulation
Package (VASP)\cite{Kresse1993_11169} was employed.
The DFT calculations were
carried out using projector-augmented wave\cite{Blochl1994_17953}
pseudopotentials\cite{Kresse1999_1758} which
are based on PBE\cite{Perdew1996_3865} all-electron atomic calculations, and a
plane wave basis set. An energy cutoff of 1000~eV
(73.50~Ry or 36.75~$E_{\text{h}}$) was used together
with an electronic SCF tolerance of 10$^{-7}$~eV. For geometry
optimizations, forces were converged such that the
maximum force per atom is at most 10$^{-2}$~eV/\AA.
Note that in the ADS41 set (see SI, Tab.~ST25),
\{H$_{2}$O,CH$_{3}$OH\}@Pt111 using SCAN\cite{Sun2015_036402}
and \{H$_{2}$O,CH$_{3}$OH,C$_{3}$H$_{8}$,C$_{4}$H$_{10}$\}@Pt111
using \mbox{SCAN-rVV10}\cite{Peng2016_041005} required a looser criterion of \mbox{$2\cdot10^{-2}$~eV/\AA} \ because of
numerical instabilities, which are absent in r$^{2}$SCAN.\cite{Furness2020_8208} The
k-point spacing for the surface (in 2 dimensions)
and bulk (in 3 dimensions) calculations was at
most 0.018~\AA$^{-1}$. A Gaussian smearing of KS occupation numbers with
a width of $2\cdot10^{-2}$~eV was used for surfaces
and bulk systems.

For data sets involving molecules, the unit cell has been
prepared as follows. Around the largest system in
the set, a box was constructed such that there is
at least 20~\AA \ in between periodic images in all
directions. This box was employed for all molecules
in the corresponding data set. For atoms, a cell of
23~\AA \ was used.

To compare the performance of the \mbox{VCML-rVV10}
functional, calculations were also performed with
PBE,\cite{Perdew1996_3865} PBE-D3,\cite{Perdew1996_3865,Grimme2010_154104} MS2,\cite{Sun2013_044113}
SCAN,\cite{Sun2015_036402} r$^{2}$SCAN,\cite{Furness2020_8208}
\mbox{SCAN-rVV10},\cite{Peng2016_041005} MCML,\cite{Brown2021_2004} and
\mbox{MCML-rVV10} (introduced in this work). For the error analysis,
the mean error (ME) and mean absolute error (MAE) are computed
\begin{align}
    \text{ME} & = \frac{1}{N} \sum_{i=1}^{N} x_{i}^{\text{calc}} - x_{i}^{\text{ref}} \\
   \text{MAE} & = \frac{1}{N} \sum_{i=1}^{N} \left| x_{i}^{\text{calc}} - x_{i}^{\text{ref}} \right|,
\end{align}
where $x_{i}^{\text{calc}}$ is the calculated value
and $x_{i}^{\text{ref}}$ the corresponding reference value.
These values are provided in the SI, Tab.~ST7, 9, 11-20, 22-25 and 27.
All individual errors for any system of a data set are
computed by $x_{i}^{\text{calc}} - x_{i}^{\text{ref}}$.

\section{Benchmark data sets}
\label{sec:datasets}
To obtain a multi-purpose XC
functional from the proposed methodology, see \secref{fitting},
a variety of training data is required.
Our training data consists of molecular,
surface as well as bulk properties, as described below.
The systems of each data set and all calculated
properties are provided in the SI, Tab.~ST6-27.

\paragraph{The DBH24 set\cite{Zheng2007_569}}
                 consists of 12 forward and 12 reverse
                 reaction barrier heights of small
                 molecules. Reference values are the
                 best estimates, consisting of quantum chemical
                 and experimental data, according to \citetdot{Zheng2009_808}
                 Total energies were computed on the
                 reference structures; no structural
                 optimizations were carried out.
\paragraph{The RE42 set\cite{Wellendorf2012_235149}}
                 contains 42 reaction energies involving
                 45 molecules of the G2/97\cite{Curtiss1997_1063}
                 test set. Experimental reference values are
                 taken from \citetdot{Wellendorf2012_235149} The
                 geometries of all molecules were fully optimized
                 before their total energy was used to calculate
                 the reaction energies.
\paragraph{The S66x8 set\cite{Rezac2011_2427}} has 66
                 molecular complexes at eight distances,
                 the interaction of which is dominated by
                 non-covalent bonding. Reference values
                 are the recommended dissociation energies, calculated
                 by MP2-F12 energies at the basis set limit combined
                 with CCSD(Tc$_{\text{sc}}$)-F12 correlation,
                 according to \citetdot{Brauer2016_20905}
                 Total energies were computed at the
                 reference geometries, thus no geometry
                 optimizations were carried out.
\paragraph{The W4-11 set\cite{karton2011w4}} contains 140
                 atomization energies of small to medium sized
                 molecules. Reference values are taken from \citetcomma{karton2011w4} which are
                 based on the Weizmann-4 (W4) computational thermochemistry method.
                 No structural optimizations were carried out,
                 and total energies were computed on the reference
                 structures.
\paragraph{The SOL62 set} is based on a set of 64 solids
                 proposed by \citetdot{Zhang2018_063020} Of that
                 set we removed all elements heavier than Au, as
                 for those systems (Pb, Th) the predicted errors
                 will typically be very large
                 due to strong relativistic effects.\cite{Wellendorf2012_235149}
                 The reference values are taken from the supplemental
                 information of \citetcomma{Zhang2018_063020} where
                 zero-point energy corrections of PBE were subtracted
                 from the experimental values. All unit cells were fully
                 optimized. Afterwards, the lattice constants $a_{\text{lat}}$ and bulk
                 moduli $B$ were computed using an equation of state (EOS)
                 fit\cite{Alchagirov2001_224115} from five energies
                 around and at the minimum volume. Further, cohesive
                 energies $E_{\text{coh}}$ were computed with respect to the isolated
                 atoms in their energetically preferred magnetic
                 ground state. This data set is split into the three
                 subset $a_{\text{lat}}$@SOL62, $E_{\text{coh}}$@SOL62, and $B$@SOL62.
\paragraph{The ADS41 set\cite{Wellendorff2015_36,Sharada2019_035439}}
                 consists of 41 surface reaction energies. Experimental
                 reference values are taken from the collection in \citetdot{Sharada2019_035439}
                 The atomic positions of the isolated molecules, the
                 isolated surfaces as well as the combined systems were
                 optimized. For the clean surfaces and the combined systems,
                 the cell was constrained to the bulk lattice dimensions,
                 while introducing at least 15~\AA~of vacuum in the
                 out-of-plane direction. The vacuum layer is the same
                 for the isolated surface and the combined system, ensuring
                 consistent results.
                 The surfaces are constructed to have 4 layers. The lower two
                 are fixed at the bulk lattice positions, while the
                 top two layers (and the adsorbed molecule) are optimized.
                 Adsorption energies are computed per
                 adsorbate, thus the total adsorption energy is divided
                 by the number of adsorbates in the computational cell. Because ADS41 contains
                 Co and Ru surfaces, we additionally optimized the unit cells
                 of either hcp solid to obtain the corresponding lattice
                 constants. These solids are not part of the SOL62 set, and
                 were only optimized for ADS41. The ADS41 data set is
                 subdivided into physisorption-dominated reactions (15 reaction
                 energies, denoted as $E_{\text{ads}}^{\text{phy}}$@ADS41) and chemisorption-dominated
                 reactions (26 reaction energies, denoted as  $E_{\text{ads}}^{\text{che}}$@ADS41).

\section{Functional fitting approach}
\label{sec:fitting}
\subsection{General outline}
In the following, the meta-GGA as well as the vdW
fitting procedure is presented. The procedure to
obtain the meta-GGA functional form has been outlined
in \citet{Brown2021_2004} for the MCML functional. We
will briefly discuss its main components here. Our
exchange enhancement factor is expanded in a series
of Legrendre polynomials $P$ for $s$ and $\alpha$ as
\begin{equation}
    F_{\text{X}}(s(\textbf{r}),\alpha(\textbf{r})) = \sum_{i=0}^{7} \sum_{j=0}^{7} c_{ij} P_{i}\left(\hat{s}(\textbf{\textbf{r}})\right) P_{j}\left(\hat{\alpha}(\textbf{r})\right) \label{eq:legendreexpand} ,
\end{equation}
where
\begin{equation}
    \hat{s}(\textbf{r}) = \frac{2s(\textbf{r})^{2}}{\eta+s(\textbf{r})^{2}} - 1
\label{eq:s_trans}
\end{equation}
maps the semi-infinite interval of reduced density
gradients $s$ to the interval spanned by the Legrendre
polynomials, $[-1,1]$. Here, $\eta=\kappa/\mu^{\rm GE}$
is used to represent the exchange enhancement of
PBEsol\cite{Perdew2008_136406} by the first two
Legendre polynomials $P_0\!\left(\hat s(\textbf{\textbf{r}})\right)$
and $P_1\!\left(\hat s(\textbf{\textbf{r}})\right)$.
Further, $1+\kappa=1.804$ is the local Lieb-Oxford
bound,\cite{Lieb1981_427,Perdew1991_11} and $\mu^{\rm GE}=10/81$
is the lowest-order coefficient of the gradient
correction to the free electron gas exchange energy.\cite{Antoniewicz1985_6779}
Similarly,
\begin{equation}
    \hat{\alpha}(\textbf{r}) = \frac{(1-\alpha(\textbf{r})^{2})^3}{1+\alpha(\textbf{r})^{3} + 4\alpha(\textbf{r})^{6}}
\end{equation}
maps the semi-infinite interval of $\alpha$ to the
interval $[-0.25,1]$. This form of $\hat{\alpha}$
coincides with the GGA-weighting function of the MS2 functional.\cite{Sun2013_044113}
The transformations of $s$ and $\alpha$ are visualized in the SI, Fig.~SF1.
Optimizing the 64 coefficients $c_{ij}$ in \eqref{legendreexpand}, as described in the next sections,
delivers an optimal semi-local meta-GGA functional. 
The physical constraints introduced for MCML are
also applied to our functional, \textit{i.e.}, the LDA limit
with $F_{\text{X}} = 1$ for $s=0$ and $\alpha=1$,
the exchange gradient expansion for $s \approx 0$
at $\alpha=1$, and the cancellation of the spurious
Hartree energy in the H atom for $\alpha = 0$.
The LDA limit is enforced via
\begin{equation}
    \sum_{i=0}^7\sum_{k=0}^3 (-1)^{i+k} \frac{(2k)!}{2^{2k}(k!)^2} \, c_{i,2k} = 1 \label{eq:ldalimit},
\end{equation}
while the exchange gradient expansion is described with a curvature of $2\mu^{\rm GE}$ for $s \approx 0$ at $\alpha=1$
\begin{equation}
    \sum_{i=0}^7\sum_{k=0}^3 -(-1)^{i+k} \frac{(2k)!}{2^{2k}(k!)^2}  \frac{2i(i+1)}{\eta} \, c_{i,2k} = 2\mu^{\rm GE} \label{eq:experturb}.
\end{equation}
The spurious Hartree energy of the hydrogen atom of $\frac{5}{16}$~$E_{\text{h}}$ is cancelled with
\begin{equation}
    \left(\frac{162}{\pi^{2}}\right)^{\frac{1}{3}} \sum_{i=0}^7 \sum_{j=0}^7 \int\limits_0^\infty {\rm d}r\, r^2 y^{2} P_i\big(\hat s^{\text{H}}\big) \, c_{ij} = \frac{5}{16},
 \label{eq:hydrogenconstraint_new} \end{equation}
where $y = \exp(-4r/3)$, $\hat s^{\text{H}} = 2p/[\eta+p] - 1$,
and $p = s^{2} = (6\pi)^{-2/3}/y$
is the square of the reduced density gradient of the $1s$ hydrogen atom ground state.

Besides PBE, the functionals considered in this study fulfill all three constraints \eqsref{ldalimit}{experturb}{hydrogenconstraint_new}.

\subsection{Regularized optimization of $F_{\text{X}}$}
We extend the approach introduced in \citet{Brown2021_2004} by including
the vdW description via rVV10.
The vdW description is optimized based
on the functional form, after which our exchange functional
is optimized based on a given vdW parameterization.
MCML is used a starting point for the optimization.
We took the product of the non-self-consistent (nonSCF)
MAEs for all data sets as the cost function for the optimization. These
nonSCF predictions are explained in \citetcomma{Brown2021_2004}
and are briefly outlined in the SI, Sec.~3. The cost function
\begin{equation}
    \theta = \prod_{i} \left(\text{MAE}_{i}\right)^{w_{i}}
\label{eq:proderror}
\end{equation}
was minimized using the Nelder-Mead simplex
algorithm,\cite{Nelder1965_308,Gao2012_259,Virtanen2020_261}
raising each MAE to a weight $w_{i}$. Adjusting the weights
results in different fits, see \secref{weights}. This enables
tuning trade-offs in performance for different data sets against each other.

\subsection{Choosing weights}
\label{sec:weights}
\begin{figure}[h]
    \centering
    \includegraphics[width=0.5\textwidth]{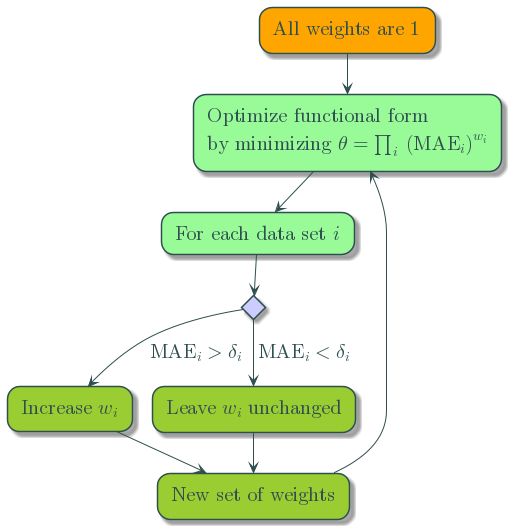}
    \caption{Adjustment of the weights in the functional
    optimization, see \eqref{proderror}. Here, $\delta_{i}$
    are thresholds for each data set, while $w_{i}$ are
    the corresponding weights. The evaluation was done
    100 times and the best weights, \textit{i.e.}, the best compromise
    for all nonSCF MAEs, was taken for further evaluation.}
    \label{fig:weights}
\end{figure}

The weights were chosen according to \figref{weights}.
As a starting point, all weights were set to~1. The
functional form is optimized, and the nonSCF predictions
of the MAE for each data set are computed. If an MAE is
beyond a predefined threshold, the weight for the data
set is increased (see SI, Tab.~ST1 for values). After this modification,
the functional is optimized once again and the process starts anew. This optimization route
was performed 100 times. The best possible
weights, \textit{i.e.}, the best compromise between all predicted
MAEs, were adjusted to minimize the errors
further. The final weights (see SI, Tab.~ST2) were then taken to obtain
the optimal $c_{ij}$ (see SI, Tab.~ST3).

\subsection{Enforce smoothness}
\begin{figure}[h!!]
    \centering
    \includegraphics[width=0.49\textwidth]{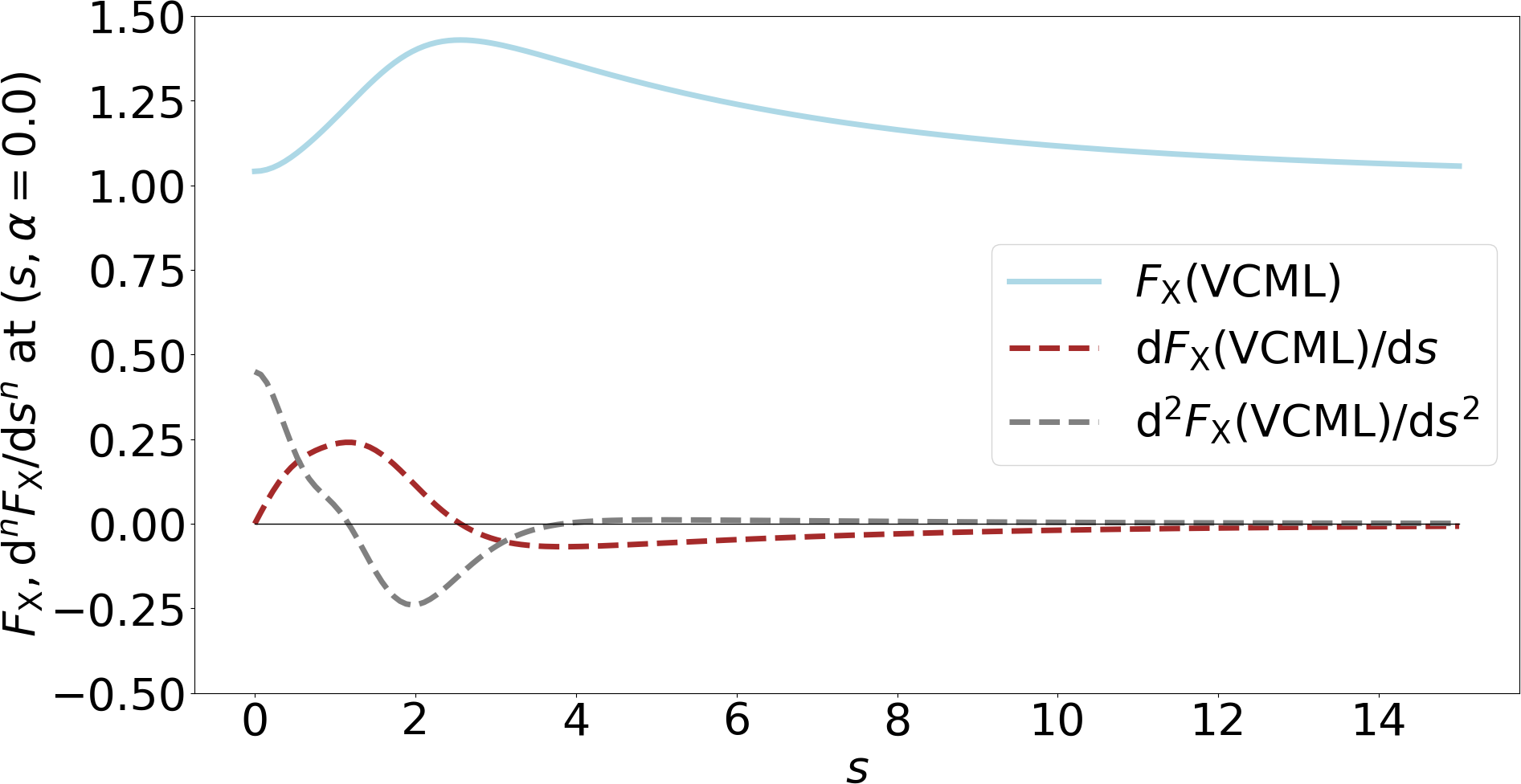}
    \caption{$F_{\text{X}}$ as well as the first and
    second derivative of $F_{\text{X}}$ with respect
    to the reduced density gradient $s$ at $\alpha=0$
    for VCML, which is the semilocal exchange part of
    \mbox{VCML-rVV10}. While $\text{d}F_{\text{X}}/\text{d}s$
    shows exactly one sign change,
    $\text{d}^{2}F_{\text{X}}/\text{d}s^{2}$ shows two.
    As such, the regularizations regarding the sign changes of derivatives of $F_{\text{X}}$ work as intended.
    This enforces a certain smoothness of the functional.}
    \label{fig:derivatives}
\end{figure}
To avoid oscillatory behavior of high-order polynomial fits, the $F_{\text{X}}$
is enforced to be smooth. The number of sign changes beyond
given thresholds in the first and second derivatives of $F_{\text{X}}$
with respect to $s$ and $\alpha$ is penalized.
A penalty is added if there is more than one sign change in the first derivatives.
This allows the exchange enhancement factor to have one extremum
in $s$ for a given $\alpha$. Further, a penalty is applied if there are more than two sign changes
in the second derivatives. With that, the functional should go smoothly
towards and away from the extremum.
Derivatives of the Legendre polynomials used to expand $F_\text{X}$ are efficiently
computed via recursion.
The zeros of the resulting polynomials corresponding to the first and
second derivatives are obtained as the eigenvalues of the companion matrices.
This sign-change penalty technique enables us to enforce smoothness of $F_\text{X}$
in a computationally straightforward way.
As an example, the first and second derivative
of $F_{\text{X}}$ with respect to $s$ at $\alpha=0$ (single
orbital limit) for VCML, the semilocal exchange part of
\mbox{VCML-rVV10}, are shown in \figref{derivatives}.
As one can see, the proposed regularizations work as
intended in making the functional smooth.
Note that there are different mathematical
forms of $F_\text{X}$ not based on polynomial expansions allowing for other techniques
to enforce constraints and smoothness.\cite{Sparrow2021__}

\subsection{Summary of entire procedure to obtain VCML}
\begin{figure}[h!!!!!!!!!]
    \centering
    \includegraphics[width=0.49\textwidth]{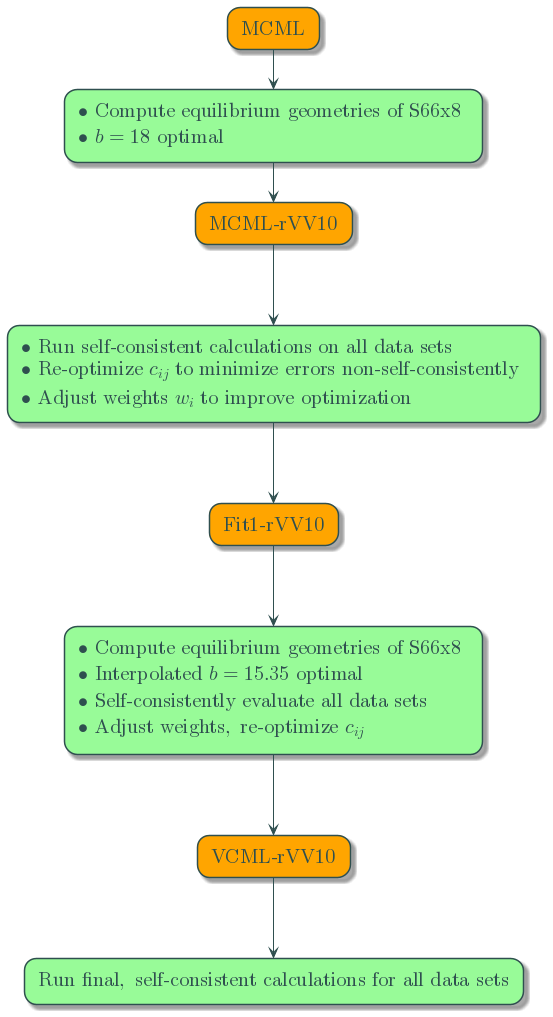}
    \caption{Procedure to arrive at the \mbox{VCML-rVV10}
    functional form. Starting from MCML, the
    optimal $b$ parameter of the rVV10 methodology is
    determined. The resulting \mbox{MCML-rVV10} functional is
    used to compute all data sets self-consistently. Based
    on these results, the functional form was adjusted
    according to the minimization of \eqref{proderror}.
    The adjustment of the weights is illustrated in \figref{weights}.
    This results in an intermediate functional form,
    called Fit1-rVV10. For this functional, the $b$ parameter
    was re-optimized, and all data sets were re-calculated
    self-consistently. Finally, the functional form was
    re-optimized once again, resulting in \mbox{VCML-rVV10}.
    Due to careful consideration of the errors in S66x8 as described in the SI, Sec.~3,
    the $b$ parameter needed no further optimization.}
    \label{fig:procedure}
\end{figure}

\figref{procedure} summarizes the fitting procedure.
It starts by taking the MCML functional. An optimal $b$ parameter for rVV10 is determined.
For this, the equilibrium structures of S66x8 and their resulting errors were used.
Various $b$ values were analyzed, ranging from 3 to 25
with a stepsize of 1. The optimal value, $b = 18$,
defines \mbox{MCML-rVV10}. After calculating all data sets
(see \secref{datasets}) with \mbox{MCML-rVV10}, the functional
form was re-optimized using the methodology outlined
above. This led to an
intermediate functional form, Fit1-rVV10.

For Fit1-rVV10,
the $b$ value was re-optimized, using a range of $b$
values between 13 and 21 and a smaller
stepsize of 0.25. From this search, an optimal $b$ value
of 15.35 was interpolated. After that, all properties of
all data sets were recalculated self-consistently. The
resulting data is used once again to re-optimize the functional
form. For this final optimization, we made sure that the
nonSCF errors for the S66x8 remain small, see SI, Sec.~3 for further details.
This avoids re-optimizing the $b$ parameter. Accordingly,
the $b$ parameter for the final \mbox{VCML-rVV10} functional
is also 15.35.

In summary, the $b$ parameter as well as the
functional form were optimized twice, starting from MCML.
Given an overall good performance of \mbox{VCML-rVV10} in comparison
to \mbox{MCML-rVV10} and other functionals (see Sec.~\ref{sec:results}),
we terminate the functional optimization loop here.

\section{Results and discussion}
\label{sec:results}

\subsection{Comparing $F_{\text{X}}$ to other functionals}
\begin{figure}[h]
    \centering
    \includegraphics[width=0.49\textwidth]{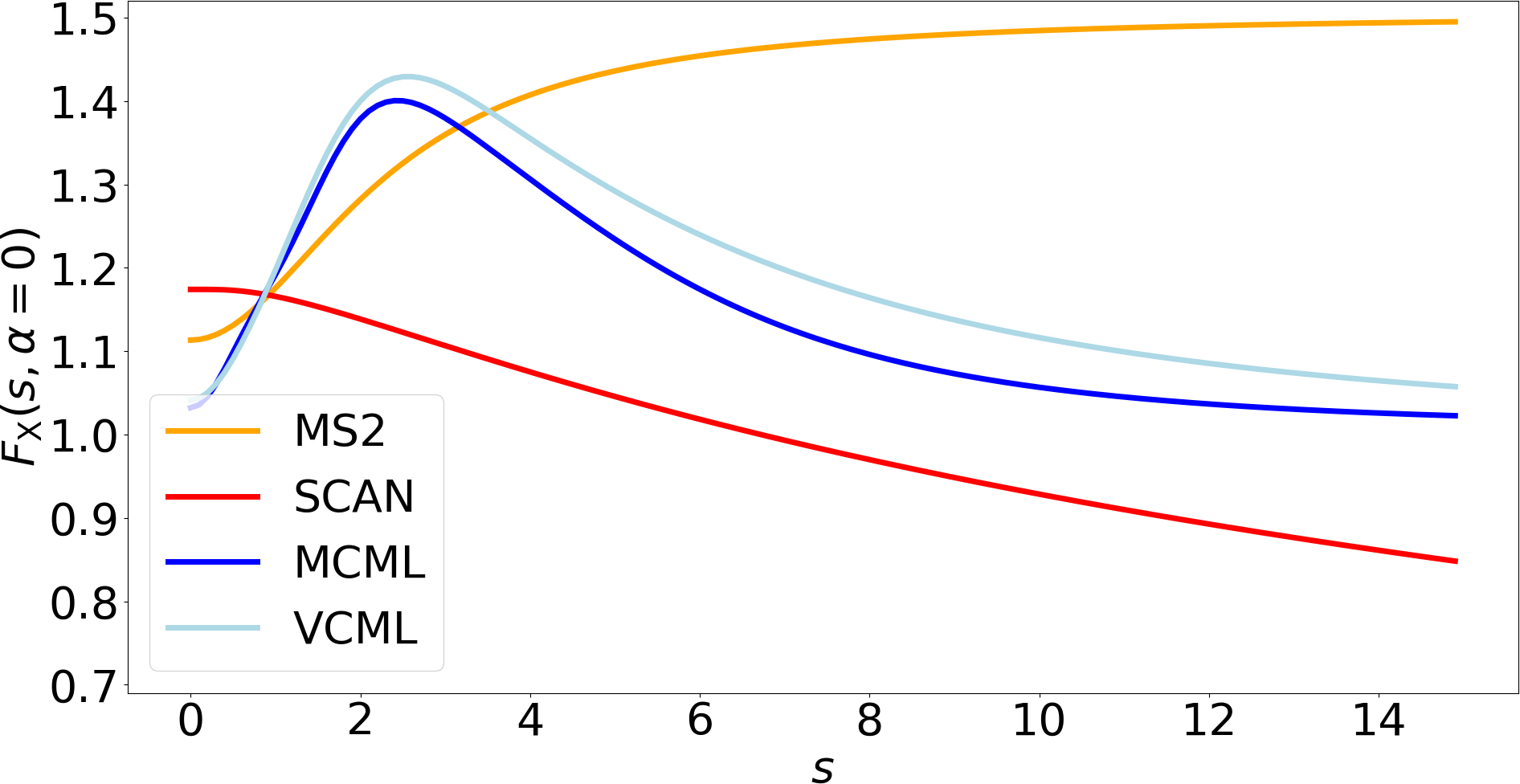}
    \includegraphics[width=0.49\textwidth]{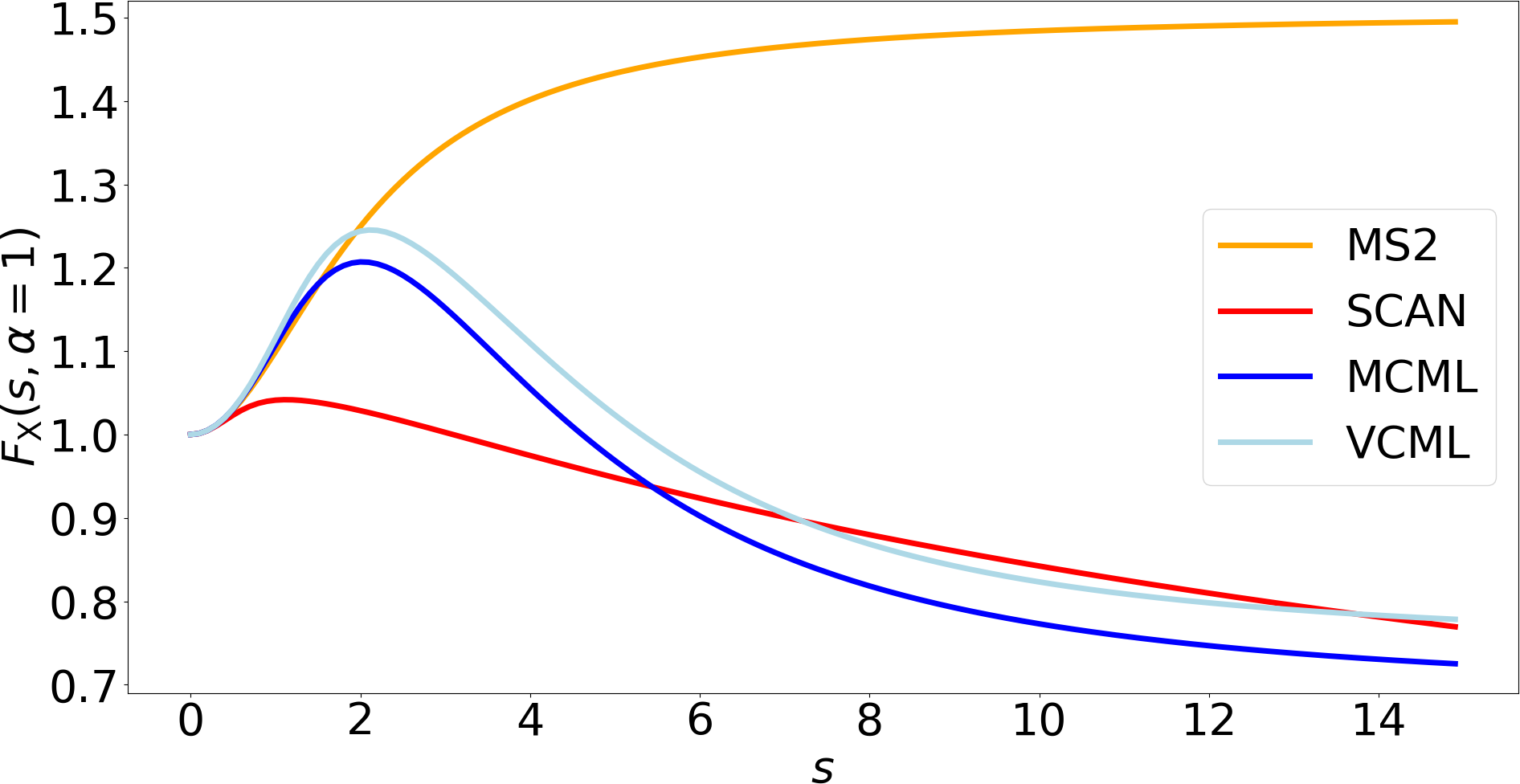}
    \includegraphics[width=0.49\textwidth]{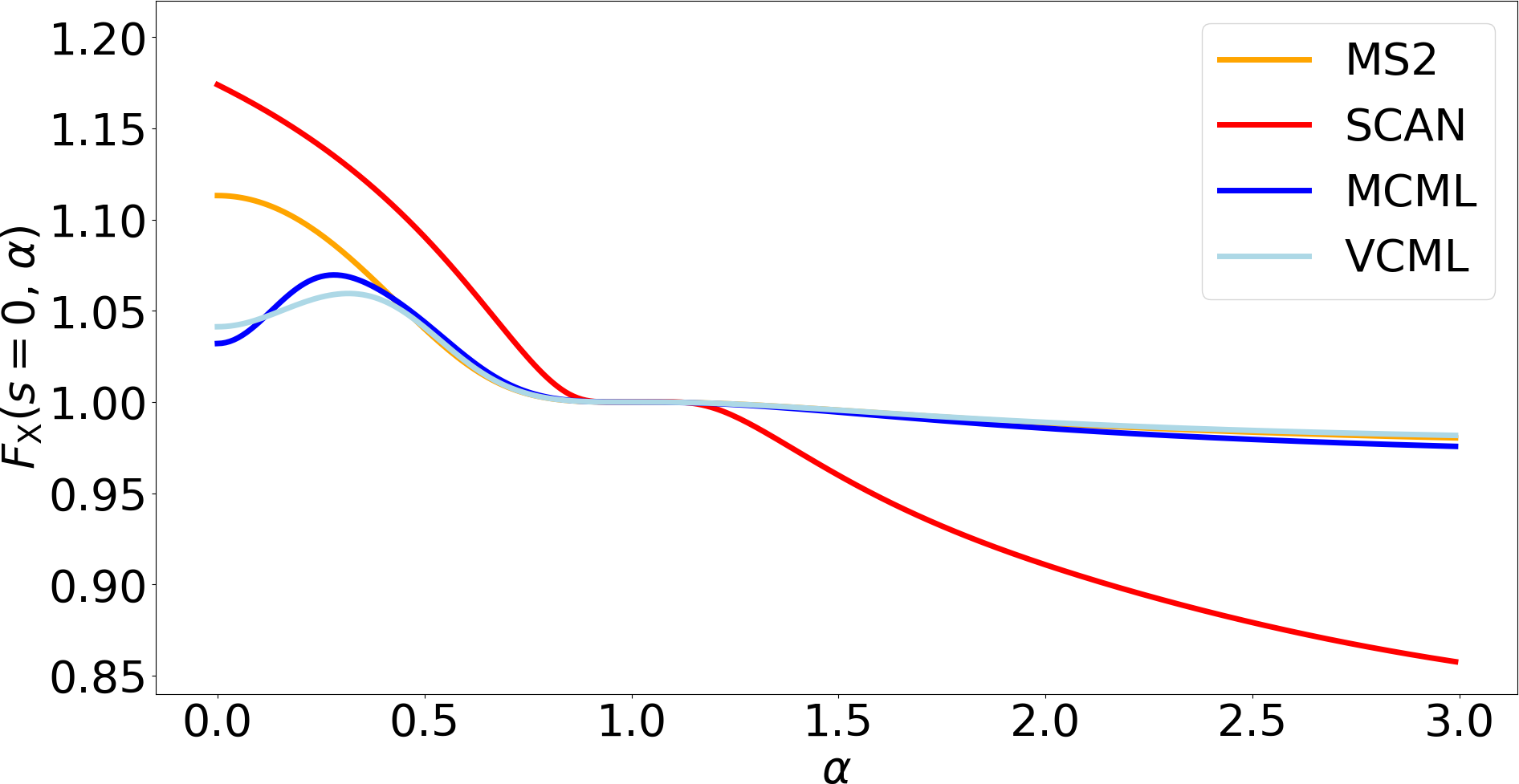}
    \caption{Exchange enhancement factor for different regimes of
    $s$ and $\alpha$ for MS2, SCAN, MCML and VCML.
    Top: $F_{\text{X}}$ for $\alpha=0$ over a range of $s$.
    Middle: $F_{\text{X}}$ for $\alpha=1$ over a range of $s$.
    Bottom: $F_{\text{X}}$ for $s=0$ over a range of $\alpha$.
    One can clearly see that $F_{\text{X}}(s=0, \alpha=1) = 1$; this is the LDA
    limit that is fulfilled by all the presented functionals,
    as is the curvature in this point given by the exchange gradient expansion.}
    \label{fig:FX}
\end{figure}

The $F_{\text{X}}$ of MS2, SCAN,
MCML as well as the new VCML are plotted in \figref{FX}. In certain regions
of $s$ and $\alpha$, VCML is similar to MS2, e.g.,
\mbox{$F_{\text{X}}(s=0, \alpha > 0.5)$}. On the other hand, VCML
is similar to SCAN for $F_{\text{X}}(s > 8, \alpha=1)$.
While the VCML and MCML exchange enhancement factor are similar
for low reduced density gradients, the VCML exchange enhancement
is markedly larger at reduced density gradients $s\gtrsim2$.
The maximum value of VCML's $F_{\text{X}}$ is 1.432, which is well below
the Lieb-Oxford bound of 1.804.\cite{Perdew2014_18A533} Note that remaining
below this bound was not explicitly enforced, nor was the decaying behavior with increasing $s$.

\subsection{Performance on data sets}
\begin{figure*}[t]
    \centering
    \includegraphics[width=0.99\textwidth]{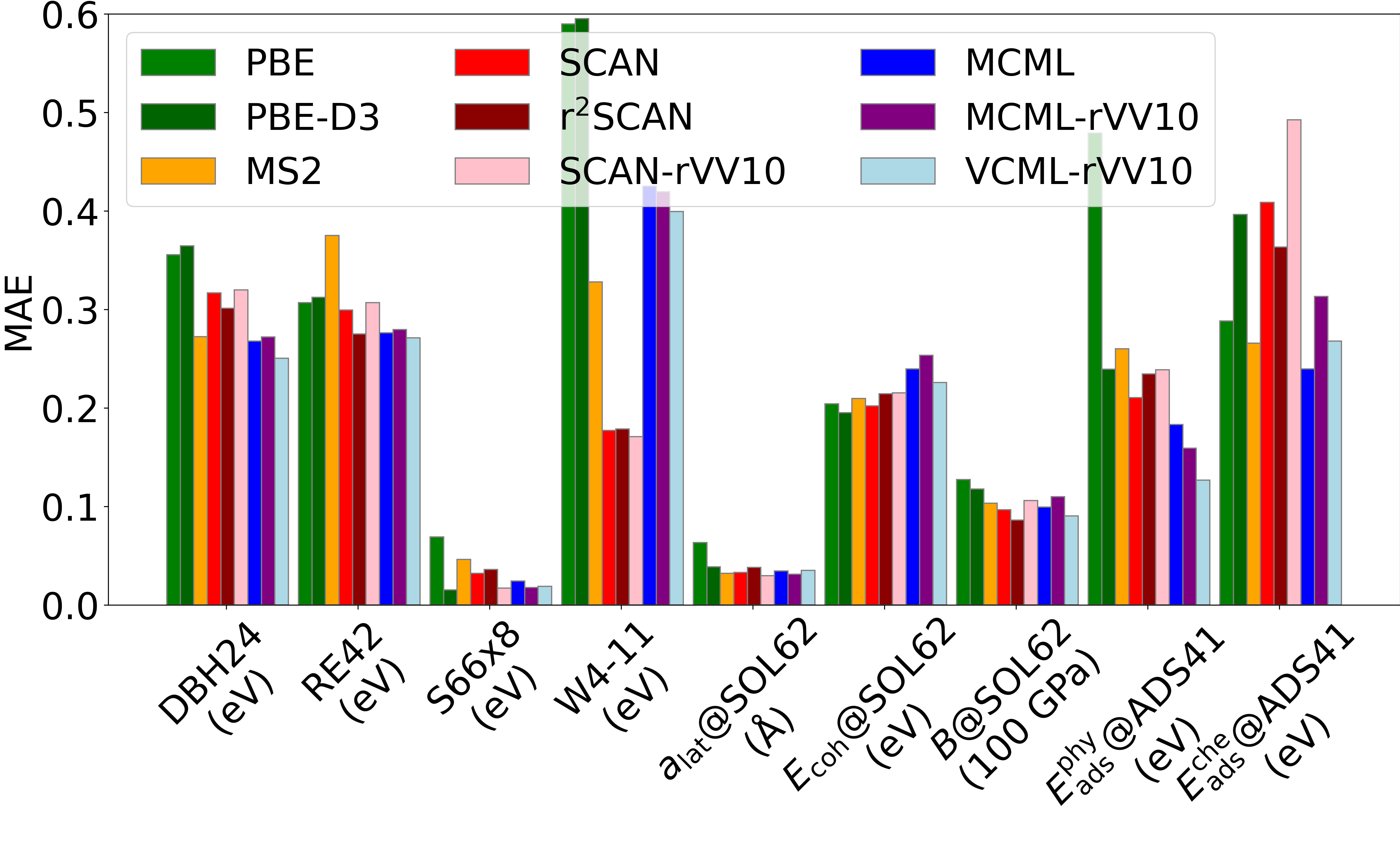}
    \caption{Comparison of the mean absolute errors (MAEs) in the
    different data sets for all employed functionals.
    The units per data set are provided. 
    Detailed plots with all errors per data set of all functionals can be found in the SI, Fig.~SF2-13.}
    \label{fig:compMAE}
\end{figure*}

A summary of the MAEs for all
data sets using the functionals described in the main
text is shown in \figref{compMAE}. For all values of MEs and MAEs see SI, Tab.~ST4 and ST5.

Overall, our \mbox{VCML-rVV10} functional shows small
errors for a range of physical properties.
This confirms that we have created a multi-purpose functional. It
should be noted that our fitting approach favors errors
of certain data sets over others. There is always a
compromise between making a specific error better,
while making some other error(s) worse. Accordingly,
the improvements seen for \mbox{VCML-rVV10} over, e.g., MCML
is the best overall compromise we found.
For a more detailed analysis of the resulting errors, see below.
\paragraph{For DBH24} \mbox{VCML-rVV10} performs best, followed
                 by MCML. Barrier heights are poorly described in
                 DFT approximations based on semi-local exchange;
                 all functionals produce a relatively
                 large error for the barriers due to large
                 self-interaction error in the regime of stretched bonds.\cite{Shahi2019_174102}
\paragraph{For RE42} again \mbox{VCML-rVV10} performs best,
                 followed by r$^{2}$SCAN. Thus, the description of
                 the reaction between molecules is described well by \mbox{VCML-rVV10}.
\paragraph{For S66x8} \mbox{SCAN-rVV10}, \mbox{MCML-rVV10} and
                 \mbox{VCML-rVV10} perform similarly, and clearly outperform any
                 functional that does not include any vdW correction. This
                 is a proof of concept that the rVV10 methodology works as
                 intended. It furthermore shows that refitting the functional
                 form from MCML to VCML did not diminish the prediction of these
                 non-covalent interactions.
\paragraph{For W4-11} we find a poor performance of \mbox{VCML-rVV10}. As noted
                 in \citetcomma{Brown2021_2004} MCML performs worse on atomization
                 energies than other tested functionals. This is confirmed as
                 seen in \figref{compMAE}. With \mbox{VCML-rVV10}, we actually improve the MAE by about 20~meV.
                 However, we find there is a trade-off especially between $E_{\textrm{ads}}^{\textrm{che}}$@ADS41
                 and $E_{\textrm{coh}}$@SOL62. Atomization energies of solids and molecules cannot be
                 improved further here without deteriorating the description of surface chemistry.
\paragraph{For SOL62} \mbox{VCML-rVV10} performs very similar for $a_{\text{lat}}$@SOL62 compared to the other \mbox{meta-GGAs}.
                 All of them perform well on lattice constants,
                 with differences in their MAE of at most 0.005~\AA.

                 Furthermore, \mbox{VCML-rVV10} outperforms
                 MCML as well as \mbox{MCML-rVV10} for $E_{\text{coh}}$@SOL62.
                 MCML is not an optimal functional for the description
                 of cohesive energies;\cite{Brown2021_2004} adding
                 rVV10 on-top makes matters worse. Due to the fact
                 that we reshaped the functional form, we arrive at
                 an MAE which is only about 10~meV worse than
                 \mbox{SCAN-rVV10}, but already 15~meV better than MCML.

                 For $B$@SOL62, \mbox{VCML-rVV10} is the second best functional
                 after r$^{2}$SCAN, clearly outperforming the other functionals
                 that employ vdW-corrections in the form of rVV10.

                 Overall, \mbox{VCML-rVV10} performs well for
                 solids regarding the three properties
                 studied here. We do not diminish the
                 performance on the solid properties and
                 thus maintain the multi-purpose character of \mbox{VCML-rVV10}.
\paragraph{For ADS41} we obtain the smallest
                 errors in the physisorption-dominated systems
                 using \mbox{VCML-rVV10}, outperforming \mbox{MCML-rVV10} and
                 \mbox{SCAN-rVV10}. All functionals without rVV10
                 have, unsurprisingly, larger errors.

                 For chemisorption-dominated systems, \mbox{VCML-rVV10}
                 performs significantly better than all other
                 vdW-supplemented functionals considered here;
                 the resulting MAE is a good as MS2. As such,
                 \mbox{VCML-rVV10} is the second best functional (together
                 with MS2) for these systems. Only MCML gives smaller
                 errors, at the price of not explicitly accounting for dispersion forces.

For all data sets it is clear that only adding rVV10 on-top of
MCML, i.e., \mbox{MCML-rVV10}, does not provide a good vdW-meta-GGA functional.
Reshaping the functional form based on a given vdW correction
drastically improves the performance of the functional, making it
applicable to a wide range of physical properties while also
treating non-local interactions via the rVV10 methodology.

\subsection{Graphene on Ni}
\begin{figure}[h]
    \centering
    \includegraphics[width=0.49\textwidth]{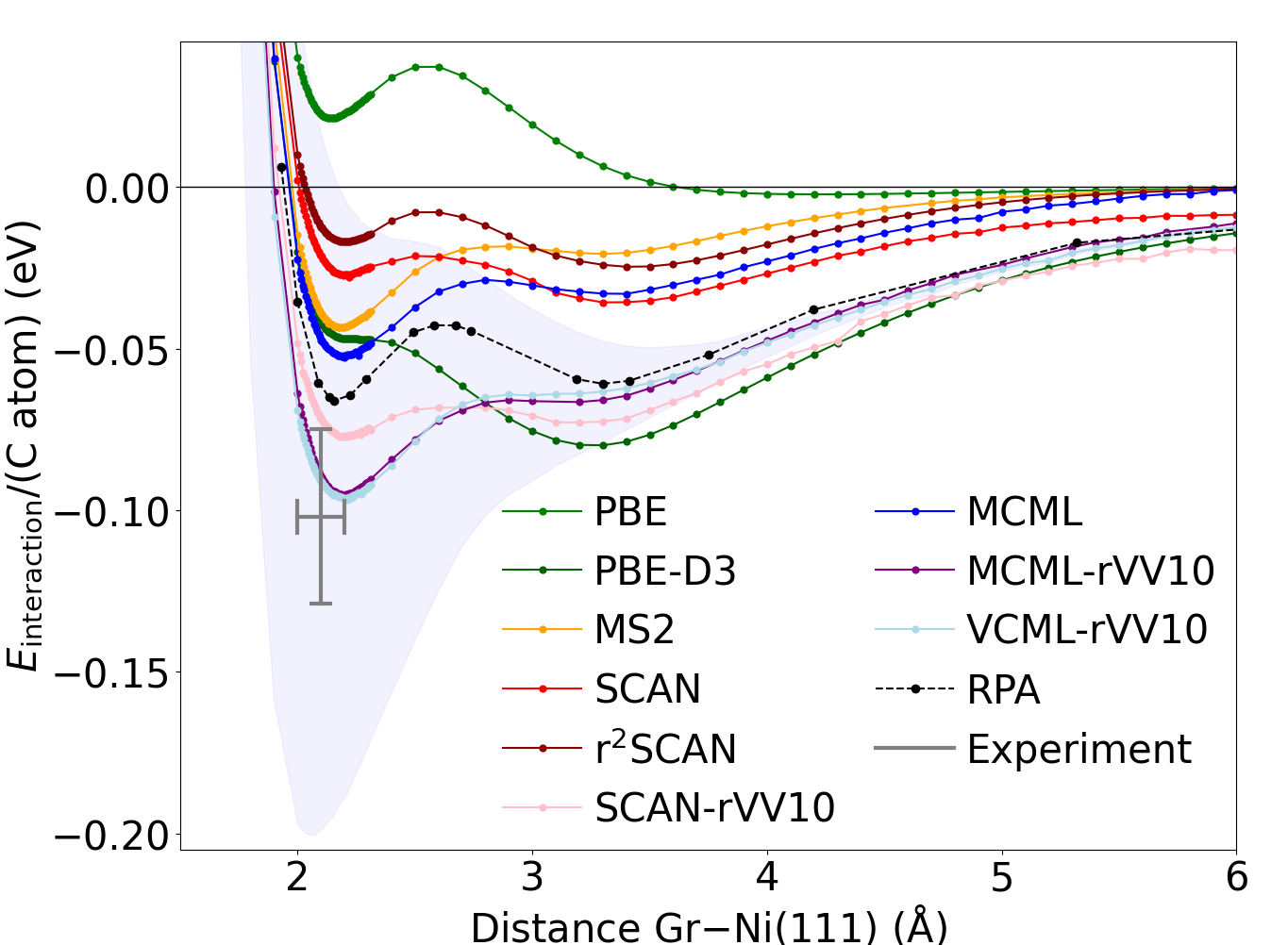}
    \caption{Interaction energy of graphene on Ni(111)
    at several distances using different functionals.
    Only \mbox{MCML-rVV10} and \mbox{VCML-rVV10} capture the experimental value of the first
    minimum\cite{Shepard2019_154702} correctly. \mbox{VCML-rVV10}
    is also energetically very close to the RPA\cite{Mittendorfer2011_201401} for the
    second, physisorption-dominated minimum.
    The shaded area indicates $\pm 1$ standard deviation
    calculated from a Bayesian ensemble of 5000 perturbations
    to the VCML exchange enhancement.}
    \label{fig:GrNi}
\end{figure}

To test \mbox{VCML-rVV10} outside of the training data shown
in \figref{compMAE}, we calculated the interaction energy
of graphene on a Ni(111) surface. Previous theoretical investigations
found that there likely are two distinct minima.\cite{Olsen2013_075111,Shepard2019_154702}
These minima can be characterized as a chemisorption
minimum at a smaller distance between graphene and Ni,
and a physisorption minimum at a larger distance. In
experiments, usually only the first minimum is found.\cite{Shepard2019_154702}
In the literature, reference calculations have been
carried out using the random phase approximation (RPA).\cite{Mittendorfer2011_201401}
It has been noted that the RPA is likely to underestimate
the chemisorption minimum.\cite{Olsen2013_075111,Shepard2019_154702}
This can be seen when comparing to experimental
estimates,\cite{Shepard2019_154702} which are shown as the grey cross in \figref{GrNi}.
The second minimum, on the other hand, is likely
captured well by the RPA.

In \figref{GrNi}, we compare
the interaction energy per C atom calculated with
all functionals considered in this article. As a note,
the optimal lattice constant for Ni for each functional
was employed. Of all the functionals, only \mbox{MCML-rVV10}
and \mbox{VCML-rVV10} describe the energy of the first minimum
accurately, with \mbox{VCML-rVV10} being within 6~meV of the average
estimate of the experimental chemisorption energy.\cite{Shepard2019_154702}
All other functionals underestimate this
minimum or predict it to be as strongly bound as the physisorbed case.
Further, \mbox{VCML-rVV10} is energetically closest
to RPA for the second minimum.
As a note, \mbox{VCML-rVV10} and \mbox{MCML-rVV10} are the only functionals
that are very close, within 5~meV, to
the RPA values at larger distances. As such, \mbox{VCML-rVV10} describes both
chemisorption and physi\-sorption accurately, indicating a well-balanced functional.


The decomposition of $E_\text{X}$ into contributions from the different
Legendre polynomial products in the expansion of the exchange-enhancement
factor $F_\text{X}$ allows for efficient non-self-consistent estimates of
changes in $E_\text{X}$ due to perturbations of $F_\text{X}$. Following the
strategy outlined in Refs.~\onlinecite{Brown2021_2004} and \onlinecite{Mortensen2005_216401},
we use such perturbations to estimate the uncertainty in the interaction energy
of graphene with Ni(111) in the example above using Bayesian inference.
Perturbations to $F_\text{X}$ are drawn randomly with a probability
$\propto \exp(-\text{Loss}_2(\theta)/\tau)$. $\text{Loss}_2(\theta)$ is the squared error
with respect to the reference fitting data as a function of the parameters $\theta$
defining $F_\text{X}$ under fulfillment of the constraints \eqsref{ldalimit}{experturb}{hydrogenconstraint_new}.
For the computation of $\text{Loss}_2(\theta)$, the energies in each data set
are normalized such that the MAE of each data set is one. The fictitious
temperature $\tau$ is determined such that the residual fitting error of
\mbox{VCML-rVV10} is reproduced (see Ref.~\onlinecite{Brown2021_2004} for further
details on the implementation of the error estimates). The predicted uncertainty
for the interaction of graphene with Ni(111) is shown as the shaded area in
Fig.~\ref{fig:GrNi}. The Bayesian error estimate for the chemisorption energy is about 0.2~eV,
which is consistent with the residual error on the chemisorption benchmark data, see \figref{compMAE}.
The ensemble error covers all stable adsorption minima.
The predicted error vanishes at larger distances, where \mbox{VCML-rVV10} and RPA agree very well.

\section{Conclusion}
\label{sec:conclusion}
In this work, a new meta-GGA van der Waals (vdW) exchange-correlation (XC) functional
called \mbox{VCML-rVV10} is introduced. This functional is obtained from
a simultaneous optimization of both semi-local exchange via a multi-parameter
model and non-local correlation of rVV10-type. \mbox{VCML-rVV10} was trained on a
number of data sets representing several chemical and physical properties.
This enables the multi-purpose character of the functional. In addition,
several constraints of the exact XC functional were enforced for physicality of
the \mbox{VCML-rVV10} model.

The newly introduced functional shows good performance in comparison to
all other functionals tested, some of which also employing the rVV10
methodology. It also performs very well on the description of graphene
on Ni(111), which was not included in the training data. This benchmark
system requires a balanced, accurate description of surface physisorption as well as chemisorption.

In the future, the introduced methodology of simultaneous, regularized
optimization of several parts of an XC functional could be extended to
hybrid functionals, where a to-be-fitted amount of screened exact
exchange would be admixed to the to-be-optimized density functional.

\section{Acknowledgement}
This research was supported by the U.S.\ Department of Energy, Office of Science, Office of Basic Energy Sciences,
Chemical Sciences, Geosciences, and Biosciences Division, Catalysis Science Program to the SUNCAT Center for Interface Science and Catalysis.
We greatly appreciate the help of James Furness for providing a patch for VASP making r$^{2}$SCAN available.

\section{Supporting Information}
\begin{itemize}[noitemsep]
    \item Further details for rVV10
    \item Visualization of transformation of $s$ and $\alpha$
    \item Details of the functional fitting procedure
    \item Fitting weights
    \item Final coefficients $c_{ij}$
    \item Tables for mean and mean absolute errors
    \item Tables with all calculated values
    \item Figures with all evaluated errors
\end{itemize}

\section{Data availability statement}
All data, including the VASP inputs/outputs as well as all calculated
values, is available at GitLab \url{https://gitlab.com/kaitrepte/vcml_data}.
The ADS41 adsorbate system structures can also be found on
catalysis-hub,\cite{Winther2019_1} see \url{https://www.catalysis-hub.org/publications/KaiData-driven2022}.
VCML-rVV10 is available in \textsc{libXC},\cite{Lehtola2018_1} see \url{https://github.com/ElectronicStructureLibrary/libxc}.
An alternative FORTRAN routine for evaluation of the VCML exchange energy can be found at \url{https://github.com/vossgroup/CML}.

\FloatBarrier

\bibliography{kai}

\begin{thebibliography}{55}
\providecommand{\natexlab}[1]{#1}
\providecommand{\url}[1]{\texttt{#1}}
\expandafter\ifx\csname urlstyle\endcsname\relax
  \providecommand{\doi}[1]{doi: #1}\else
  \providecommand{\doi}{doi: \begingroup \urlstyle{rm}\Url}\fi

\bibitem[Hohenberg and Kohn(1964)]{Hohenberg1964_B864}
P.~Hohenberg and W.~Kohn.
\newblock {Inhomogeneous Electron Gas}.
\newblock \emph{Phys. Rev.}, 136:\penalty0 B864--B871, 1964.
\newblock \doi{10.1103/PhysRev.136.B864}.

\bibitem[Kohn and Sham(1965)]{Kohn1965_A1133}
W.~Kohn and L.~J. Sham.
\newblock {Self-Consistent Equations Including Exchange and Correlation
  Effects}.
\newblock \emph{Phys. Rev.}, 140:\penalty0 A1133, 1965.
\newblock \doi{10.1103/PhysRev.140.A1133}.

\bibitem[Kohn et~al.(1996)Kohn, Becke, and Parr]{Kohn1996_12974}
W.~Kohn, A.~D. Becke, and R.~G. Parr.
\newblock {Density Functional Theory of Electronic Structure}.
\newblock \emph{J. Phys. Chem.}, 100:\penalty0 12974, 1996.
\newblock \doi{10.1021/jp960669l}.

\bibitem[Burke(2012)]{Burke2011_150901}
K.~Burke.
\newblock {Perspective on density functional theory}.
\newblock \emph{J. Chem. Phys.}, 136:\penalty0 150901, 2012.
\newblock \doi{10.1063/1.4704546}.

\bibitem[Mardirossian and Head-Gordon(2017)]{Mardirossian2017_2315}
N.~Mardirossian and M.~Head-Gordon.
\newblock {Thirty years of density functional theory in computational
  chemistry: an overview and extensive assessment of 200 density functionals}.
\newblock \emph{Mol. Phys.}, 115:\penalty0 2315--2372, 2017.
\newblock \doi{10.1080/00268976.2017.1333644}.

\bibitem[Becke(1993)]{Becke1993_5648}
A.~D. Becke.
\newblock {Density‐functional thermochemistry. III. The role of exact
  exchange}.
\newblock \emph{J. Chem. Phys.}, 98:\penalty0 5648--5652, 1993.
\newblock \doi{10.1063/1.464913}.

\bibitem[Zhao et~al.(2005)Zhao, Schultz, and Truhlar]{Zhao2005_161103}
Y.~Zhao, N.~E. Schultz, and D.~G. Truhlar.
\newblock {Exchange-correlation functional with broad accuracy for metallic and
  nonmetallic compounds, kinetics, and noncovalent interactions}.
\newblock \emph{J. Chem. Phys.}, 123:\penalty0 161103, 2005.
\newblock \doi{10.1063/1.2126975}.

\bibitem[Peverati and Truhlar(2014)]{Perverati2011_20120476}
R.~Peverati and D.~G. Truhlar.
\newblock {Quest for a universal density functional: the accuracy of density
  functionals across a broad spectrum of databases in chemistry and physics}.
\newblock \emph{Philos. Trans. R. Soc. A}, 372:\penalty0 20120476, 2014.
\newblock \doi{10.1098/rsta.2012.0476}.

\bibitem[Mardirossian and Head-Gordon(2014)]{Mardirossian2014_9904}
N.~Mardirossian and M.~Head-Gordon.
\newblock {$\omega$B97X-V: A 10-parameter{,} range-separated hybrid{,}
  generalized gradient approximation density functional with nonlocal
  correlation{,} designed by a survival-of-the-fittest strategy}.
\newblock \emph{Phys. Chem. Chem. Phys.}, 16:\penalty0 9904--9924, 2014.
\newblock \doi{10.1039/C3CP54374A}.

\bibitem[Perdew et~al.(1996)Perdew, Burke, and Ernzerhof]{Perdew1996_3865}
J.~P. Perdew, K.~Burke, and M.~Ernzerhof.
\newblock {Generalized Gradient Approximation Made Simple}.
\newblock \emph{Phys. Rev. Lett.}, 77:\penalty0 3865--3868, 1996.
\newblock \doi{10.1103/PhysRevLett.77.3865}.

\bibitem[Perdew et~al.(2008)Perdew, Ruzsinszky, Csonka, Vydrov, Scuseria,
  Constantin, Zhou, and Burke]{Perdew2008_136406}
J.~P. Perdew, A.~Ruzsinszky, G.~I. Csonka, O.~A. Vydrov, G.~E. Scuseria, L.~A.
  Constantin, X.~Zhou, and K.~Burke.
\newblock {Restoring the Density-Gradient Expansion for Exchange in Solids and
  Surfaces}.
\newblock \emph{Phys. Rev. Lett.}, 100:\penalty0 136406, 2008.
\newblock \doi{10.1103/PhysRevLett.100.136406}.

\bibitem[Sun et~al.(2015)Sun, Ruzsinszky, and Perdew]{Sun2015_036402}
J.~Sun, A.~Ruzsinszky, and J.~P. Perdew.
\newblock {Strongly Constrained and Appropriately Normed Semilocal Density
  Functional}.
\newblock \emph{Phys. Rev. Lett.}, 115:\penalty0 036402, 2015.
\newblock \doi{10.1103/PhysRevLett.115.036402}.

\bibitem[Garza et~al.(2018)Garza, Bell, and Head-Gordon]{Garza2018_3083}
A.~J. Garza, A.~T. Bell, and M.~Head-Gordon.
\newblock {Nonempirical Meta-Generalized Gradient Approximations for Modeling
  Chemisorption at Metal Surfaces}.
\newblock \emph{J. Chem. Theory Comput.}, 14:\penalty0 3083--3090, 2018.
\newblock \doi{10.1021/acs.jctc.8b00288}.

\bibitem[Furness et~al.(2020)Furness, Kaplan, Ning, Perdew, and
  Sun]{Furness2020_8208}
J.~W. Furness, A.~D. Kaplan, J.~Ning, J.~P. Perdew, and J.~Sun.
\newblock {Accurate and Numerically Efficient r$^2$SCAN Meta-Generalized
  Gradient Approximation}.
\newblock \emph{J. Phys. Chem. Lett.}, 11:\penalty0 8208--8215, 2020.
\newblock \doi{10.1021/acs.jpclett.0c02405}.

\bibitem[Perdew et~al.(2014)Perdew, Ruzsinszky, Sun, and
  Burke]{Perdew2014_18A533}
J.~P. Perdew, A.~Ruzsinszky, J.~Sun, and K.~Burke.
\newblock {Gedanken densities and exact constraints in density functional
  theory}.
\newblock \emph{J. Chem. Phys.}, 140:\penalty0 18A533, 2014.
\newblock \doi{10.1063/1.4870763}.

\bibitem[Bokdam et~al.(2017)Bokdam, Lahnsteiner, Ramberger, Sch\"afer, and
  Kresse]{Bokdam2017_145501}
M.~Bokdam, J.~Lahnsteiner, B.~Ramberger, T.~Sch\"afer, and G.~Kresse.
\newblock {Assessing Density Functionals Using Many Body Theory for Hybrid
  Perovskites}.
\newblock \emph{Phys. Rev. Lett.}, 119:\penalty0 145501, 2017.
\newblock \doi{10.1103/PhysRevLett.119.145501}.

\bibitem[Zhang et~al.(2018)Zhang, Reilly, Tkatchenko, and
  Scheffler]{Zhang2018_063020}
G.-X. Zhang, A.~M. Reilly, A.~Tkatchenko, and M.~Scheffler.
\newblock Performance of various density-functional approximations for cohesive
  properties of 64 bulk solids.
\newblock \emph{New J. Phys.}, 20:\penalty0 063020, 2018.
\newblock \doi{10.1088/1367-2630/aac7f0}.

\bibitem[Brown et~al.(2021)Brown, Maimaiti, Trepte, Bligaard, and
  Voss]{Brown2021_2004}
K.~Brown, Y.~Maimaiti, K.~Trepte, T.~Bligaard, and J.~Voss.
\newblock {MCML: Combining physical constraints with experimental data for a
  multi-purpose meta-generalized gradient approximation}.
\newblock \emph{J. Comput. Chem.}, 42:\penalty0 2004--2013, 2021.
\newblock \doi{https://doi.org/10.1002/jcc.26732}.

\bibitem[Sparrow et~al.(2021)Sparrow, Ernst, Quady, and \mbox{DiStasio}
  Jr.]{Sparrow2021__}
Z.~M. Sparrow, B.~G. Ernst, T.~K. Quady, and R.~A. \mbox{DiStasio} Jr.
\newblock {CASE21: Uniting Non-Empirical and Semi-Empirical Density Functional
  Approximation Strategies using Constraint-Based Regularization}.
\newblock \emph{arXiv}, 2109.12560, 2021.

\bibitem[Zhao et~al.(2006)Zhao, Schultz, and Truhlar]{Zhao2006_364}
Y.~Zhao, N.~E. Schultz, and D.~G. Truhlar.
\newblock {Design of Density Functionals by Combining the Method of Constraint
  Satisfaction with Parametrization for Thermochemistry, Thermochemical
  Kinetics, and Noncovalent Interactions}.
\newblock \emph{J. Chem. Theory Comput.}, 2:\penalty0 364--382, 2006.
\newblock \doi{10.1021/ct0502763}.

\bibitem[Duanmu and Truhlar(2017)]{Duanmu2017_835}
K.~Duanmu and D.~G. Truhlar.
\newblock {Validation of Density Functionals for Adsorption Energies on
  Transition Metal Surfaces}.
\newblock \emph{J. Chem. Theory Comput.}, 13:\penalty0 835--842, 2017.
\newblock \doi{10.1021/acs.jctc.6b01156}.

\bibitem[Sharada et~al.(2019)Sharada, Karlsson, Maimaiti, Voss, and
  Bligaard]{Sharada2019_035439}
S.~M. Sharada, R.~K.~B. Karlsson, Y.~Maimaiti, J.~Voss, and T.~Bligaard.
\newblock Adsorption on transition metal surfaces: {Transferability} and
  accuracy of {DFT} using the {ADS41} dataset.
\newblock \emph{Phys. Rev. B}, 100:\penalty0 035439, 2019.
\newblock \doi{10.1103/PhysRevB.100.035439}.

\bibitem[Wellendorff et~al.(2012)Wellendorff, Lundgaard, M\o{}gelh\o{}j,
  Petzold, Landis, N\o{}rskov, Bligaard, and Jacobsen]{Wellendorf2012_235149}
J.~Wellendorff, K.~T. Lundgaard, A.~M\o{}gelh\o{}j, V.~Petzold, D.~D. Landis,
  J.~K. N\o{}rskov, T.~Bligaard, and K.~W. Jacobsen.
\newblock {Density functionals for surface science: Exchange-correlation model
  development with Bayesian error estimation}.
\newblock \emph{Phys. Rev. B}, 85:\penalty0 235149, 2012.
\newblock \doi{10.1103/PhysRevB.85.235149}.

\bibitem[Klime{\v{s}} et~al.(2009)Klime{\v{s}}, Bowler, and
  Michaelides]{Klimes2009_02201}
J.~Klime{\v{s}}, D.~R. Bowler, and A.~Michaelides.
\newblock {Chemical accuracy for the van der Waals density functional}.
\newblock \emph{J. Phys.: Condens. Matter}, 22:\penalty0 022201, 2009.
\newblock \doi{10.1088/0953-8984/22/2/022201}.

\bibitem[Vydrov and Van~Voorhis(2010)]{Vydrov2010_244103}
O.~A. Vydrov and T.~Van~Voorhis.
\newblock {Nonlocal van der Waals density functional: The simpler the better}.
\newblock \emph{J. Chem. Phys.}, 133:\penalty0 244103, 2010.
\newblock \doi{10.1063/1.3521275}.

\bibitem[Sabatini et~al.(2013)Sabatini, Gorni, and
  de~Gironcoli]{Sabatini2013_041108}
R.~Sabatini, T.~Gorni, and S.~de~Gironcoli.
\newblock {Nonlocal van der Waals density functional made simple and
  efficient}.
\newblock \emph{Phys. Rev. B}, 87:\penalty0 041108, 2013.
\newblock \doi{10.1103/PhysRevB.87.041108}.

\bibitem[Peng et~al.(2016)Peng, Yang, Perdew, and Sun]{Peng2016_041005}
H.~Peng, Z.-H. Yang, J.~P. Perdew, and J.~Sun.
\newblock {Versatile van der Waals Density Functional Based on a
  Meta-Generalized Gradient Approximation}.
\newblock \emph{Phys. Rev. X}, 6:\penalty0 041005, 2016.
\newblock \doi{10.1103/PhysRevX.6.041005}.

\bibitem[Perdew et~al.(2009)Perdew, Ruzsinszky, Csonka, Constantin, and
  Sun]{Perdew2009_026403}
J.~P. Perdew, A.~Ruzsinszky, G.~I. Csonka, L.~A. Constantin, and J.~Sun.
\newblock {Workhorse Semilocal Density Functional for Condensed Matter Physics
  and Quantum Chemistry}.
\newblock \emph{Phys. Rev. Lett.}, 103:\penalty0 026403, 2009.
\newblock \doi{10.1103/PhysRevLett.103.026403}.

\bibitem[Perdew and Schmidt(2001)]{Perdew2001_1}
J.~P. Perdew and K.~Schmidt.
\newblock Jacob’s ladder of density functional approximations for the
  exchange-correlation energy.
\newblock \emph{AIP Conf. Proc.}, 577:\penalty0 1--20, 2001.
\newblock \doi{10.1063/1.1390175}.

\bibitem[Kresse and Furthm{\"u}ller(1993)]{Kresse1993_11169}
G.~Kresse and J.~Furthm{\"u}ller.
\newblock Efficient iterative schemes for ab initio total-energy calculations
  using a plane-wave basis set.
\newblock \emph{Phys. Rev. B}, 54:\penalty0 11169, 1993.
\newblock \doi{10.1103/PhysRevB.54.11169}.

\bibitem[Bl{\"o}chl(1994)]{Blochl1994_17953}
P.~E. Bl{\"o}chl.
\newblock {Projector augmented-wave method}.
\newblock \emph{Phys. Rev. B}, 50:\penalty0 17953, 1994.
\newblock \doi{10.1103/PhysRevB.50.17953}.

\bibitem[Kresse and Joubert(1999)]{Kresse1999_1758}
G.~Kresse and D.~Joubert.
\newblock From ultrasoft pseudopotentials to the projector augmented-wave
  method.
\newblock \emph{Phys. Rev. B}, 59:\penalty0 1758, 1999.
\newblock \doi{10.1103/PhysRevB.59.1758}.

\bibitem[Grimme et~al.(2010)Grimme, Antony, Ehrlich, and
  Krieg]{Grimme2010_154104}
S.~Grimme, J.~Antony, S.~Ehrlich, and H.~Krieg.
\newblock {A consistent and accurate ab initio parametrization of density
  functional dispersion correction (DFT-D) for the 94 elements H-Pu}.
\newblock \emph{J. Chem. Phys.}, 132:\penalty0 154104, 2010.
\newblock \doi{10.1063/1.3382344}.

\bibitem[Sun et~al.(2013)Sun, Haunschild, Xiao, Bulik, Scuseria, and
  Perdew]{Sun2013_044113}
J.~Sun, R.~Haunschild, B.~Xiao, I.~W. Bulik, G.~E. Scuseria, and J.~P. Perdew.
\newblock {Semilocal and hybrid meta-generalized gradient approximations based
  on the understanding of the kinetic-energy-density dependence}.
\newblock \emph{J. Chem. Phys.}, 138:\penalty0 044113, 2013.
\newblock \doi{10.1063/1.4789414}.

\bibitem[Zheng et~al.(2007)Zheng, Zhao, and Truhlar]{Zheng2007_569}
J.~Zheng, Y.~Zhao, and D.~G. Truhlar.
\newblock {Representative Benchmark Suites for Barrier Heights of Diverse
  Reaction Types and Assessment of Electronic Structure Methods for
  Thermochemical Kinetics}.
\newblock \emph{J. Chem. Theory Comput.}, 3:\penalty0 569, 2007.
\newblock \doi{10.1021/ct600281g}.

\bibitem[Zheng et~al.(2009)Zheng, Zhao, and Truhlar]{Zheng2009_808}
J.~Zheng, Y.~Zhao, and D.~G. Truhlar.
\newblock {The DBH24/08 Database and Its Use to Assess Electronic Structure
  Model Chemistries for Chemical Reaction Barrier Heights}.
\newblock \emph{J. Chem. Theory Comput.}, 5:\penalty0 808--821, 2009.
\newblock \doi{10.1021/ct800568m}.

\bibitem[Curtiss et~al.(1997)Curtiss, Raghavachari, Redfern, and
  Pople]{Curtiss1997_1063}
L.~A. Curtiss, K.~Raghavachari, P.~C. Redfern, and J.~A. Pople.
\newblock {Assessment of {Gaussian-2} and density functional theories for the
  computation of enthalpies of formation}.
\newblock \emph{J. Chem. Phys.}, 106:\penalty0 1063--1079, 1997.
\newblock \doi{10.1063/1.473182}.

\bibitem[Rezac et~al.(2011)Rezac, Riley, and Hobza]{Rezac2011_2427}
J.~Rezac, K.~E. Riley, and P.~Hobza.
\newblock {S66: A Well-balanced Database of Benchmark Interaction Energies
  Relevant to Biomolecular Structures}.
\newblock \emph{J. Chem. Theory Comput.}, 7:\penalty0 2427--2438, 2011.
\newblock \doi{10.1021/ct2002946}.

\bibitem[Brauer et~al.(2016)Brauer, Kesharwani, Kozuch, and
  Martin]{Brauer2016_20905}
B.~Brauer, M.~K. Kesharwani, S.~Kozuch, and J.~M.~L. Martin.
\newblock {The S66x8 benchmark for noncovalent interactions revisited:
  explicitly correlated ab initio methods and density functional theory}.
\newblock \emph{Phys. Chem. Chem. Phys.}, 18:\penalty0 20905--20925, 2016.
\newblock \doi{10.1039/C6CP00688D}.

\bibitem[Karton et~al.(2011)Karton, Daon, and Martin]{karton2011w4}
A.~Karton, S.~Daon, and J.~M.~L. Martin.
\newblock {W4-11: A high-confidence benchmark dataset for computational
  thermochemistry derived from first-principles W4 data}.
\newblock \emph{Chem. Phys. Lett.}, 510:\penalty0 165--178, 2011.
\newblock \doi{https://doi.org/10.1016/j.cplett.2011.05.007}.

\bibitem[Alchagirov et~al.(2001)Alchagirov, Perdew, Boettger, Albers, and
  Fiolhais]{Alchagirov2001_224115}
A.~B. Alchagirov, J.~P. Perdew, J.~C. Boettger, R.~C. Albers, and C.~Fiolhais.
\newblock {Energy and pressure versus volume: {Equations} of state motivated by
  the stabilized jellium model}.
\newblock \emph{Phys. Rev. B}, 63:\penalty0 224115, 2001.
\newblock \doi{10.1103/PhysRevB.63.224115}.

\bibitem[Wellendorff et~al.(2015)Wellendorff, Silbaugh, Garcia-Pintos,
  Nørskov, Bligaard, Studt, and Campbell]{Wellendorff2015_36}
J.~Wellendorff, T.~L. Silbaugh, D.~Garcia-Pintos, J.~K. Nørskov, T.~Bligaard,
  F.~Studt, and C.~T. Campbell.
\newblock {A benchmark database for adsorption bond energies to transition
  metal surfaces and comparison to selected DFT functionals}.
\newblock \emph{Surf. Sci.}, 640:\penalty0 36--44, 2015.
\newblock \doi{https://doi.org/10.1016/j.susc.2015.03.023}.

\bibitem[Lieb and Oxford(1981)]{Lieb1981_427}
E.~H. Lieb and S.~Oxford.
\newblock Improved lower bound on the indirect {Coulomb} energy.
\newblock \emph{Int. J. Quantum Chem.}, 19:\penalty0 427, 1981.
\newblock \doi{https://doi.org/10.1002/qua.560190306}.

\bibitem[Perdew(1991)]{Perdew1991_11}
J.~P Perdew.
\newblock Unified theory of exchange and correlation beyond the local density
  approximation.
\newblock In P.~Ziesche and H.~Eschrig, editors, \emph{Electronic Structure of
  Solids '91}, volume~17 of \emph{Physical Research}, pages 11--20, Berlin,
  1991. Akademie Verlag.

\bibitem[Antoniewicz and Kleinman(1985)]{Antoniewicz1985_6779}
P.~R. Antoniewicz and L.~Kleinman.
\newblock Kohn-{Sham} exchange potential exact to first order in
  {\ensuremath{\rho}}({K})/${\ensuremath{\rho}}_{0}$.
\newblock \emph{Phys. Rev. B}, 31:\penalty0 6779, 1985.
\newblock \doi{10.1103/PhysRevB.31.6779}.

\bibitem[Nelder and Mead(1965)]{Nelder1965_308}
J.~A. Nelder and R.~Mead.
\newblock {A Simplex Method for Function Minimization}.
\newblock \emph{Comput. J.}, 7:\penalty0 308, 1965.
\newblock \doi{10.1093/comjnl/7.4.308}.

\bibitem[Gao and Han(2012)]{Gao2012_259}
F.~Gao and L.~Han.
\newblock Implementing the {Nelder}-{Mead} simplex algorithm with adaptive
  parameters.
\newblock \emph{Comput. Optim. Appl.}, 51:\penalty0 259, 2012.
\newblock \doi{10.1007/s10589-010-9329-3}.

\bibitem[Virtanen et~al.(2020)Virtanen, Gommers, Oliphant, Haberland, Reddy,
  Cournapeau, Burovski, Peterson, Weckesser, Bright, {van der Walt}, Brett,
  Wilson, Millman, Mayorov, Nelson, Jones, Kern, Larson, Carey, Polat, Feng,
  Moore, {VanderPlas}, Laxalde, Perktold, Cimrman, Henriksen, Quintero, Harris,
  Archibald, Ribeiro, Pedregosa, {van Mulbregt}, and {SciPy 1.0
  Contributors}]{Virtanen2020_261}
P.~Virtanen, R.~Gommers, T.~E. Oliphant, M.~Haberland, T.~Reddy, D.~Cournapeau,
  E.~Burovski, P.~Peterson, W.~Weckesser, J.~Bright, S.~J. {van der Walt},
  M.~Brett, J.~Wilson, K.~J. Millman, N.~Mayorov, A.~R.~J. Nelson, E.~Jones,
  R.~Kern, E.~Larson, C.~J. Carey, \.I. Polat, Y.~Feng, E.~W. Moore,
  J.~{VanderPlas}, D.~Laxalde, J.~Perktold, R.~Cimrman, I.~Henriksen, E.~A.
  Quintero, C.~R. Harris, A.~M. Archibald, A.~H. Ribeiro, F.~Pedregosa, P.~{van
  Mulbregt}, and {SciPy 1.0 Contributors}.
\newblock {SciPy} 1.0: {Fundamental} algorithms for scientific computing in
  {Python}.
\newblock \emph{Nat. Methods}, 17:\penalty0 261, 2020.
\newblock \doi{10.1038/s41592-019-0686-2}.

\bibitem[Shahi et~al.(2019)Shahi, Bhattarai, Wagle, Santra, Schwalbe, Hahn,
  Kortus, Jackson, Peralta, Trepte, Lehtola, Nepal, Myneni, Neupane, Adhikari,
  Ruzsinszky, Yamamoto, Baruah, Zope, and Perdew]{Shahi2019_174102}
C.~Shahi, P.~Bhattarai, K.~Wagle, B.~Santra, S.~Schwalbe, T.~Hahn, J.~Kortus,
  K.~A. Jackson, J.~E. Peralta, K.~Trepte, S.~Lehtola, N.~K. Nepal, H.~Myneni,
  B.~Neupane, S.~Adhikari, A.~Ruzsinszky, Y.~Yamamoto, T.~Baruah, R.~R. Zope,
  and J.~P. Perdew.
\newblock {Stretched or noded orbital densities and self-interaction correction
  in density functional theory}.
\newblock \emph{J. Chem. Phys.}, 150:\penalty0 174102, 2019.
\newblock \doi{10.1063/1.5087065}.

\bibitem[Olsen and Thygesen(2013)]{Olsen2013_075111}
T.~Olsen and K.~S. Thygesen.
\newblock {Random phase approximation applied to solids, molecules, and
  graphene-metal interfaces: From van der Waals to covalent bonding}.
\newblock \emph{Phys. Rev. B}, 87:\penalty0 075111, 2013.
\newblock \doi{10.1103/PhysRevB.87.075111}.

\bibitem[Shepard and Smeu(2019)]{Shepard2019_154702}
S.~Shepard and M.~Smeu.
\newblock {First principles study of graphene on metals with the SCAN and
  SCAN+rVV10 functionals}.
\newblock \emph{J. Chem. Phys.}, 150:\penalty0 154702, 2019.
\newblock \doi{10.1063/1.5046855}.

\bibitem[Mittendorfer et~al.(2011)Mittendorfer, Garhofer, Redinger,
  Klime\ifmmode~\check{s}\else \v{s}\fi{}, Harl, and
  Kresse]{Mittendorfer2011_201401}
F.~Mittendorfer, A.~Garhofer, J.~Redinger, J.~Klime\ifmmode~\check{s}\else
  \v{s}\fi{}, J.~Harl, and G.~Kresse.
\newblock {Graphene on Ni(111): Strong interaction and weak adsorption}.
\newblock \emph{Phys. Rev. B}, 84:\penalty0 201401, 2011.
\newblock \doi{10.1103/PhysRevB.84.201401}.

\bibitem[Mortensen et~al.(2005)Mortensen, Kaasbjerg, Frederiksen, Nørskov,
  Sethna, and Jacobsen]{Mortensen2005_216401}
J.~J. Mortensen, K.~Kaasbjerg, S.~L. Frederiksen, J.~K. Nørskov, J.~P. Sethna,
  and K.~W. Jacobsen.
\newblock Bayesian error estimation in density-functional theory.
\newblock \emph{Phys. Rev. Lett.}, 95:\penalty0 216401, 2005.
\newblock \doi{10.1103/PhysRevLett.95.216401}.

\bibitem[Winther et~al.(2019)Winther, Hoffmann, Boes, Mamun, Bajdich, and
  Bligaard]{Winther2019_1}
K.~T. Winther, M.~J. Hoffmann, J.~R. Boes, O.~Mamun, M.~Bajdich, and
  T.~Bligaard.
\newblock Catalysis-hub. org, an open electronic structure database for surface
  reactions.
\newblock \emph{Sci. data}, 6\penalty0 (1):\penalty0 1--10, 2019.
\newblock \doi{10.1038/s41597-019-0081-y}.

\bibitem[Lehtola et~al.(2018)Lehtola, Steigemann, Oliveira, and
  Marques]{Lehtola2018_1}
S.~Lehtola, C.~Steigemann, M.~J.~T. Oliveira, and M.~A.~L. Marques.
\newblock {Recent developments in LIBXC -- A comprehensive library of
  functionals for density functional theory}.
\newblock \emph{SoftwareX}, 7:\penalty0 1, 2018.
\newblock \doi{10.1016/j.softx.2017.11.002}.

\end{thebibliography}

\end{document}